\documentstyle[12pt]{article}
\begin{document}
\parskip 10pt plus 1pt
\title{Logarithm of the scale factor as a generalised coordinate in a lagrangian 
for  dark matter and dark energy}
\author{
{\it Debashis Gangopadhyay $^{a}$ and Somnath Mukherjee $^{b}$}\\
$^{a}$ S.N.Bose National Centre For Basic Sciences\\
JD-Block,Sector-III,Salt Lake, Kolkata-700098,India\\
debashis@bose.res.in\\
$^{b}$ Department of Physics,Jadavpur University,\\
Kolkata-700032,India.\\
}
\date{}
\maketitle
\baselineskip=20pt
\begin{abstract}

A lagrangian for the $k-$ essence field is set up with canonical kinetic terms
and incorporating the scaling relation of [1]. There are two degrees of freedom , 
{\it viz.},$q(t)= ln\enskip a(t)$ ($a(t)$ is the scale factor) and the scalar field 
$\phi$, and an interaction term involving $\phi$ and $q(t)$.The Euler-Lagrange 
equations are solved for $q$ and $\phi$. Using these solutions quantities of cosmological 
interest are determined. The energy density $\rho$ has a constant component
which we identify as dark energy and a component behaving as $a^{-3}$ which we
call dark matter. The pressure $p$ is  {\it negative} for time $t\rightarrow \infty$
and the sound velocity $c_{s}^{2}={\partial p\over\partial\rho} << 1$.
When dark energy dominates ,the deceleration parameter $Q\rightarrow -1$ while in 
the matter dominated era $Q\sim {1\over 2}$. The equation of state parameter  
$w={p\over \rho}$ is shown to be consistent with  $w={p\over\rho}\sim -1$ for dark 
energy domination and during the matter dominated era we have $w\sim 0$. 
Bounds for the parameters of the theory are estimated from observational data.

Keywords: k-essence models, dark matter, dark energy

PACS No: 98.80.-k
\end{abstract}
\maketitle

{\bf 1.Introduction}

The universe consists of roughly 25 percent
dark matter, 70 percent dark energy , about 4 percent free hydrogen and helium 
with the remaining one percent consisting of stars, dust, neutrinos and heavy elements.
In [1] it was shown that it is possible to unify the dark matter and dark energy 
components into a single scalar field model with the scalar field $\phi$ having a  
non-canonical kinetic term. These scalar fields are known as $k-$essence fields.
The idea of $k-$ essence first came in models of inflation [2,3]. Subsequently $k-$essence 
fields were shown to lead to models of dark energy also [4-7]. The general form of the 
lagrangian for these $k-$essence models is assumed to be a function $F(X)$ of the 
derivatives of the field (i.e.$X=\nabla_{\mu}\phi\nabla^{\mu}\phi$) and do not 
depend explicitly on $\phi$ to start with. In [1] the evolution of $\phi$ for an 
arbitrary functional form for the lagrangian has been given in terms of an exact 
analytical solution. The solution was in the form of a general scaling relation 
between the function $F$ of the derivatives of the scalar field and the scale factor $a(t)$ 
of the Robertson-Walker metric (a similar expression was first derived in [8]).
To obtain this result the scalar field potential $V(\phi)$ was assumed to be a
constant.
In [1] 
specific forms (motivated from string theory [9,10,2,3]) for the lagrangian (or pressure) 
$p$ and $F(X)$ were assumed to show that self-consistent models can be built which account for both the 
dark matter and dark energy components. Reviews on 
dark matter and dark energy can be found in references [11, 12, 13]. Literature on  
on $k-$ essence models  are in references 
[14,15,16,17,18,19,20,21,22,23,24,25].

The motivation of the present work stems from the question whether the standard  
lagrangian formalism can be used to understand the origins of dark matter
and dark energy after preserving the scaling relation in [1]. By standard formalism 
we mean that the kinetic terms corresponding to fields in the lagrangian should be canonical. 
Given the fact that the constituents of dark matter and dark energy are unknown to start with, 
it is extremely difficult to write down some sort of 
lagrangian. Yet there exists schemes (like the one described above) where a lagrangian 
can be written down although the kinetic terms are non-canonical. The problem
with such lagrangians are that one cannot use the well established methods 
of the lagrangian formalism.Moreover, if $\phi$ is a quantum field then  
studying such fields 
outside the gambit of the 
lagrangian formalism is problematic.

The basic results of this work can be summarised as follows.
Using the zero-zero component of Einstein's field equations 
and incorporating the scaling relation of [1], 
an expression for the lagrangian for the $k-$essence field is obtained.
This lagrangian has a non-canonical kinetic term. We now 
convert this  lagrangian into one with canonical kinetic terms after a redefinition of the 
variables.  
There are two degrees of freedom, {\it viz.}  
$q(t)= ln\enskip a(t)$ and $\phi$.
Note that $\dot q(t)$ is nothing but the Hubble parameter.
The resulting lagrangian is in standard form of canonical kinetic terms corresponding 
to $q(t)$ and a complicated interaction term involving the scalar field $\phi$ and $q(t)$. 
We solve
the Euler-Lagrange equations for $q$ and $\phi$. The solutions
give realistic cosmological scenarios in the context of dark matter
and dark energy. The energy density $\rho$ has a  
dark energy component and  a dark matter component. 
The pressure $p$ 
is {\it negative} for time $t\rightarrow \infty$ 
and the sound velocity $c_{s}^{2}={\partial p\over\partial\rho} << 1$.
When dark energy dominates ,i.e. $t\rightarrow\infty$,
the deceleration parameter $Q\rightarrow -1$ while in the matter dominated 
era $Q\sim {1\over 2}$. When the dark energy dominates, the equation of 
state parameter  $w={p\over \rho}$ is shown to 
be consistent with  $w={p\over\rho}\sim -1$ and during the matter dominated era
we have $w\sim 0$. Bounds can be estimated for the constants of integration in the 
theory from observational data.

{\bf 2.The Lagrangian and the solutions of the Euler-Lagrange equations}

The lagrangian for the $k-$essence field is taken as 
$$ L= -V(\phi) F(X)\eqno(1)$$
$$ X=\nabla_{\mu}\phi\nabla^{\mu}\phi\eqno(2)$$
The pressure $p$ is taken to be given by $(1)$ and the energy density given by
$$\rho = V(\phi)[ F(X) -2 X F_{X}]\eqno(3)$$
with $F_{X}\equiv {dF\over dX}$.The equation of state parameter 
$$w=p/\rho={F\over 2XF_{X}-F}\eqno(4)$$
and we take the standard expression for the the sound velocity as 
$$c_{s}^{2}={\partial p\over\partial\rho}\eqno(5)$$
For a flat Robertson Walker metric the equation for the $k-$essence field is
$$(F_{X}+2XF_{XX})\ddot\phi+3HF_{X}\dot\phi+(2XF_{X}-F){V_{\phi}\over V}=0\eqno(6)$$
$V_{\phi}\equiv {dV\over d\phi}$, $H={\dot a(t)\over a(t)}$ is the Hubble parameter.
In [1],  $V(\phi)$ was  a constant so that the third term in $(6)$
was absent and the following scaling law was obtained: 
$$X F_{X}^{2}=Ca^{-6}\eqno(7)$$
where $a$ is the scale factor and $C$ a constant. We assume  
that $V(\phi)$ is not a constant but  ${V_{\phi}\over V}$ can be made 
sufficiently small so that the third term in $(6)$ is still negligible 
and thus $(7)$ still holds (we will show explicitly that this is possible in our 
approach, refer to the discussion after equation $(38)$). Using $(7)$ and 
the zero-zero component of Einstein's field equations an expression for the lagrangian 
for the $k-$essence field is obtained as described below.
We take the Robertson-Walker metric :
$$ds^{2}= c^{2}dt^{2} - a^{2}(t)[{dr^{2}\over (1-kr^{2})} + r^{2}(d\theta^{2} + sin^{2}\theta d\phi^{2})]\eqno(8)$$
where $k=0, 1\enskip or -1$ is the curvature constant. 
The zero-zero component of Einstein's equation reads:
$$R_{00} - {1\over 2}g_{00}R = - k T_{00}\eqno(9)$$
This gives with the metric $(8)$ 
$${k\over a^{2}} + H^{2} = {8\pi G\over 3}\rho\eqno(10)$$
Using $(1),(2),(3),(9)$ and $(10)$,and for $k=0$, we arrive at
$$X F_{X}= {1\over 2X}[ F - ({3\over 8\pi G V(\phi)})H^{2}]\eqno(11)$$
Using $(7)$ to eliminate $F_{X}$  gives  
$$F(X) = 2{\sqrt C}{\sqrt X}a^{-3} + 3 {H^{2}\over 8\pi G V(\phi)}\eqno(12)$$
So the  expression for the lagrangian is obtained as 
$$L=-2{\sqrt C}{\sqrt X}a^{-3}V(\phi) - ({3\over 8\pi G}) H^{2}$$
$$=-2{\sqrt C}\sqrt {\dot\phi ^{2}-(\nabla\phi)^{2}}a^{-3}V(\phi) 
- 3{H^{2}\over 8\pi G}\eqno(13)$$
where $X=\nabla_{\mu}\phi\nabla^{\mu}\phi$.
Homogeneity and isotropy 
of spacetime imply $\phi(t,\bf x)=\phi(t)$. 
Then $(13)$ becomes
$$L=-c_{1}\dot q^{2} - c_{2} V(\phi)\dot \phi e^{-3q}\eqno(14)$$
where $q(t)=ln\enskip a(t)$, $c_{1}= 3(8\pi G)^{-1}$, $c_{2}=2 \sqrt C$.
Thus the non-standard lagrangian in $(1)$ has now been cast into a standard
form . The new lagrangian has two generalised coordinates $q(t)$ and $\phi(t)$.
$q$ has a standard kinetic term while $\phi$ does not have a kinetic part.
There is a complicated polynomial interaction betwen $q$ and $\phi$ and $\phi$ 
occurs purely through this interaction term.
The two Euler-Lagrange equations corresponding to $q(t)$ and $\phi(t)$
are respectively:
$$ {d\over dt}(2c_{1}\dot q(t))=-3 c_{2} V(\phi)\dot\phi  e^{-3 q(t)}\eqno(15)$$
$$\dot\phi = - {3 V(\phi)\over {\partial V\over \partial\phi}} \dot q\eqno(16)$$
Equation $(15)$ and $(16)$ may be looked upon as describing the 
evolution of the scale factor of the  universe. Substituting $(16)$ in $(15)$ gives 
$$ {d\over dt}(\dot q) = -8\pi G{\sqrt C}{d\over dt}(e^{-3q}){V^{2}\over V_{\phi}}\eqno(17)$$
To solve this ,let ${V^{2}\over V_{\phi}}=-A_{1}$ where $A_{1}$ is a positive constant
and $V_{\phi}={dV\over d\phi}$.This assumption means 
$$V(\phi)={A_{1}\over \phi + A_{2}}\eqno(18)$$ 
where $A_{2}$ is a constant. We shall show that 
using the solutions to the equations of motion for $q$ and $\phi$ ,~
${V_{\phi}\over V}$ ($=-{1\over \phi + A_{2}}$)
can be made as small as we please so that the third term in $(6)$ is ignorable 
and the scaling relation $(7)$ can be made to remain valid.
So $(17)$ becomes 
$$ {d\over dt}(\dot q) = 8\pi G{\sqrt C}A_{1}{d\over dt}(e^{-3q})\eqno(19)$$
which gives after one integration 
$$ \dot q = 8\pi G{\sqrt C}A_{1}(e^{-3q})+ A_{3}\eqno(20)$$
Assumption here is that $H=\dot q={\dot a\over a}\not=0$ for
$q\rightarrow\infty$. Consequently $A_{3}\not=0$. Solving
$(20)$ gives  
$$a_{c}(t)=[A_{4} e^{3A_{3}}t - {8\pi G{\sqrt C}A_{1}\over A_{3}}]^{{1\over 3}}$$
The subscript $c$ in $a_{c}$ means that this is a solution of the classical equations of motion. 
We choose $A_{4}=1$ and write $\alpha=3A_{3}$ and 
$\beta= {8\pi G{\sqrt C}A_{1}\over A_{3}}={24\pi G{\sqrt c}A_{1}\over \alpha}$.
We have chosen one constant of integration in the solution for $\phi$ to be unity 
and  $\alpha,\beta,A_{1}, A_{2}$ to be all positive.
Then the solutions for $a$, $\phi$,
$H$ and $V(\phi)$ are
$$a_{c}(t)=[ e^{\alpha t} - \beta]^{{1\over 3}}\eqno(21a)$$
$$\phi_{c}(t)=  e^{\alpha t} - \beta - A_{2}\eqno(21b)$$
$$H_{c}={\dot a_{c}\over a_{c}}= {\alpha e^{\alpha t}\over 3(e^{\alpha t} -\beta)}\eqno(21c)$$
$$V_{c}(\phi)=V(\phi_{c})=V(t)={A_{1}\over\phi_{c} + A_{2}}={A_{1}\over e^{\alpha t} - \beta}\eqno(21d)$$
We shall now calculate all cosmological quantities using these solutions only. 

{\bf 3. The energy density and the pressure}

From $(12)$, $F-2XF_{X}= {3H^{2}\over 8\pi GV}$. Hence  
$$\rho = V (F-2XF_{X}) = {3 H^{2}\over 8\pi G}\eqno(22)$$
$$p(\equiv L)= -V F = -{3 H^{2}\over 8\pi G} - 2{\sqrt C}{\sqrt X}Va^{-3}\eqno(23)$$
Using the solutions $(21)$ give
$$\rho_{c}\enskip\enskip =\enskip\enskip {3 H_{c}^{2}\over 8\pi G}\enskip\enskip
={\alpha^{2}\over 24\pi G}\enskip +\enskip {\alpha^{2}\beta \over 24\pi G}\enskip a_{c}^{-3} 
\enskip\enskip +\enskip\enskip {\alpha^{2}\beta\over 24\pi G}\enskip a_{c}^{-3}(1-\beta e^{-\alpha t})^{-1}\eqno(24)$$
$$p_{c}\enskip=\enskip-\enskip {3 H_{c}^{2}\over 8\pi G}\enskip
-\enskip 2{\sqrt C}\dot\phi_{c} V_{c}a_{c}^{-3}
=\enskip-\enskip\rho_{c}\enskip 
+\enskip 2{\sqrt C}\alpha A_{1}(1-\beta e^{-\alpha t})^{-1}\enskip a_{c}^{-3}\eqno(25)$$
$\rho$ has a constant component i.e dark energy, a  component  varying as $~ a^{-3}$ i.e dark matter;
and a third component whose variation for large times is again like $a^{-3}$ i.e.dark matter.
Substituting from $(21a)$ into $(25)$ 
and taking $t\rightarrow\infty$ the pressure $p$ (i.e. the lagrangian $L$) 
is surely  {\it negative}.Thus for time scales very much larger than the matter domination 
era we have a negative pressure which may be the source for the observed acceleration of the universe. 

{\bf 4. The equation of state} 

Expressing the second term on the right-hand side of $(25)$ in terms of $\rho_{c}$ 
gives the equation of state:
$$p_{c}\enskip=\enskip \rho_{c}\enskip - \enskip {2\alpha\over {\sqrt {24\pi G}}}\enskip\rho_{c}^{1\over 2}\eqno(26)$$
The equation of state parameter 
$$w= {p_{c}\over\rho_{c}}=  1 - {2\alpha\over {\sqrt {24\pi G}} \rho_{c}^{1\over 2}}
= -1 + 2\beta e^{-\alpha t}\eqno(27)$$
Therefore when the dark energy dominates , i.e. at times $t\rightarrow\infty$, 
we have $w\approx -1$.

We show below that both $\alpha$ and $\beta$ can be estimated  
to be positive in our  scheme if we accept the current 
conjectures that the dark matter and dark energy densities were equal at a time 
one-tenth the present age of the universe ($\sim 10^{17}$ seconds) 
and that the dark energy density is roughly twice that of the 
dark matter density at present [1]. Moreover, from consistency arguments
$\beta$ will be shown to be greater than ${1\over 2}$.
It will also be shown below that
$\alpha t\sim$  very large number so that $e^{-\alpha t}$ is small and higher powers of 
of $e^{-\alpha t}$ ignorable.
From $(24)$ we get 
$$\rho_{c}\enskip\enskip 
={\alpha^{2}\over 24\pi G} + {\alpha^{2}\beta \over 12\pi G}\enskip a_{c}^{-3}
+{\alpha^{2}\over 24\pi G} a_{c}^{-3}[\beta^{2} e^{-\alpha t}+ {\beta^{3}\over 2!}e^{-2\alpha t}
+{\beta^{4}\over 3!}e^{-3\alpha t} +......]\eqno(28)$$
So for large times the third term on right hand side of $(28)$ is exponentially
damped. This allows us to write the energy density as 
$$\rho_{c}\enskip\enskip= ~~~~ \rho_{DE}~~~~+~~~~\rho_{DM}~~~~+~~~~\rho'\eqno(29a)$$
where 
$$\rho_{DE}={\alpha^{2}\over 24\pi G}~~~~~~~;~~~~~~
\rho_{DM}= {\alpha^{2}\beta\over 12\pi G}\enskip a_{c}^{-3}\eqno(29b)$$
and $\rho'$ is the part that has negligible contribution at large times,{\it viz.}
$$\rho'=
+{\alpha^{2}\over 24\pi G} a_{c}^{-3}[\beta^{2} e^{-\alpha t}+ {\beta^{3}\over 2!}e^{-2\alpha t}
+{\beta^{4}\over 3!}e^{-3\alpha t} +......]
\eqno(29c)$$

The age of the universe is $t_{0}\sim 10^{17}$ seconds and at the present epoch indications are 
that the dark energy density is greater than the dark matter density. Let 
$\rho_{DE}\approx n\rho_{DM}$ where $n > 1$. Using $(29b)$ this gives 
$$t_{0}\sim 10^{17}= A_{0}{1\over\alpha}ln~((2n+1)\beta)\eqno(30a)$$
$A_{0}$ is a phenomenological parameter to be observationally determined.
Now the dark matter and dark energy densities were equal roughly 
when the time $t_{eq}\sim 10^{16}$seconds.
Again using $(29b)$ gives 
$$t_{eq}\sim 10^{16}= A_{eq}{1\over\alpha}ln~(3\beta)\eqno(30b)$$
where $A_{eq}$ is another parameter.
Solving $(30a)$ and $(30b)$ gives 
$$\beta\sim \Biggl({(2n+1)^{A_{0}}\over 3^{10A_{eq}}}\Biggr)^{{1\over 10A_{eq}-A_{0}}}$$
$$\alpha\sim 10^{-16} {A_{0}A_{eq}\over 10A_{eq}-A_{0}}ln~[{(2n+1)\over 3^{A_{0}}}]\eqno(31)$$
So for $n > 0$ , $\beta$ is always positive.For $\alpha$ to be positive we must have 
${2n+1\over 3^{A_{0}}}>1$. Current estimates are that $n\approx 2$ [1].Then $A_{0}\geq 1$.So 
positivity of $\alpha$ and  $\beta$ is ensured. 
We  further clarify our estimate of $\beta$ as follows. Consider 
the equation of state parameter $(27)$. In the matter dominated 
era $w\sim 0$. Let the time scale be denoted by $t_{m}$. So
$$w=-1 + 2\beta e^{-\alpha t_{m}}\approx 0$$
This gives
$$ t_{m}\approx {1\over\alpha}ln~(2\beta)\eqno(32)$$
Since physical time scales cannot be negative, this means that 
$ln(2\beta) > 0$ i.e.$2\beta > 1$ or $\beta > {1\over 2}$. This is consistent
with the similar requirement from $(30b)$ ,{\it viz.}, $ln~(3\beta)>0$ i.e.
$\beta > {1\over 3}$.

{\bf 5. The sound speed and the deceleration parameter}

The (square of ) sound speed is from equation $(26)$
$$c_{s}^{2}= {\partial p_{c}\over\partial\rho_{c}}
=1 - {\alpha\over{{\sqrt {24\pi G}}\rho_{c}^{{1\over 2}}}}
=1 - {\alpha\over\sqrt {24\pi G}}{\sqrt {24\pi G}\over\alpha}\Biggl( {e^{\alpha t}-\beta\over e^{\alpha t}}\Biggr)
=\beta e^{-\alpha t}\eqno(33)$$
So $c_{s}^{2}<<1$  provided 
$e^{\alpha t}>> \beta$, i.e. $e^{\alpha t} >> {24\pi G{\sqrt C}A_{1}\over\alpha}$.
One way to  achieved this is by choosing the constant ${\sqrt C}={\alpha c' \over 24\pi G A_{1}}$
where $c'>> 1$ so that $e^{\alpha t}>> 1$. Note that $\alpha$ and $t$ are always positive. 
So the sound speed $c_{s}^{2}<<1$. Note that these results are consistent with 
our estimates obtained before.

Now consider the deceleration parameter defined by $Q=-{a\ddot a\over (\dot a)^{2}}$.
$$Q=-{a_{c}\ddot a_{c}\over (\dot a_{c})^{2}} = - 1 + 3 \beta e^{-\alpha t}\eqno(34a)$$
Therefore when dark energy dominates i.e. $t\rightarrow \infty$ we have 
$$Q _{t\rightarrow \infty}  =  -  1\eqno(34b)$$
Let us now try to determine the behaviour of $Q$ in the matter dominated era.
During the matter dominated era $Q\sim {1\over 2}$. 
Imposing this condition on equation $(34a)$ gives 
$$-1+3\beta e^{-\alpha t_{m}}\approx {1\over 2}~~i.e.~~t_{m}~~\approx ~~{1\over \alpha}~~ln~2\beta\eqno(35)$$ 

{\bf 6.Self Consistency and fixing of parameters}

Comparing $(35)$ and $(32)$ we see that we have obtained the same 
value for the time scale of the matter dominated era evaluated 
from two different view points {\it viz.} equation of state parameter $w\sim 0$
and deceleration parameter $Q\sim {1\over 2}$.
So our approach is internally self-consistent and 
we can write the following equations ( $A_{0}, A_{eq}, A_{m}$ are 
positive, phenomenological constants):
$$t_{0}= A_{0}{1\over\alpha}ln~((2n+1)\beta)\sim 10^{17}\eqno(36a)$$
$$t_{eq}= A_{eq}{1\over\alpha}ln~(3\beta)\sim 10^{16}\eqno(36b)$$
$$t_{m}=A_{m}{1\over\alpha}ln~(2\beta)\eqno(36c)$$
The thing to note here is that under current estimates [1], $n\approx 2$ and so 
$ln~[(2n+1)\beta] > ln (3\beta) > ln (2\beta)$ i.e. $t_{0} > t_{eq} > t_{m}$ 
and this is what is physically expected. So the present theory is 
again shown to be self-consistent and this consistency will place 
bounds on the relative values of the parameters $A_{0}, A_{eq}, A_{m}$.

Let us now check whether the assumptions used in obtaining equation $(28)$ 
were consistent with the other inputs. Note that for $t\rightarrow\infty$ only 
$\rho_{DE}$ and $\rho_{DM}$ survive. We now carry out the following order of 
magnitudes analysis. Consider the time set $t=(t_{m}, t_{eq}, t_{0})$. 
Then $(36)$ can be succintly written as 
$$\alpha t \sim ln~[(2n+1)\beta]\eqno(37)$$
modulo respective constants (i.e. $A_{0}, A_{eq}, A_{m}$ ). Then $n={1\over 2}$
gives the equation $(36c)$, $n=1$ gives $(36b)$ and $n=n$ means $(36a)$. 
So $(37)$ implies 
$$\beta e^{-\alpha t}= {1\over (2n+1)}
\equiv ({1\over 2};~~ or~~ {1\over 3};~~ or~~ {1\over 5})$$
i.e. $n={1\over 2}$ corresponds to the matter dominated era, $n={1}$ 
signifies the period when the dark matter and dark energy densities 
were equal and $n= 2$ characterises the present epoch when dark energy dominates.
$n$ is expected to grow larger and larger with time as the domination by dark 
energy increases. So equation $(28)$ can also be written as 
$$\rho_{c}\enskip\enskip
={\alpha^{2}\over 24\pi G} + {\alpha^{2}\beta \over 24\pi G}\enskip a_{c}^{-3}
+{\alpha^{2}\beta\over 24\pi G}\enskip a_{c}^{-3}(1+\beta e^{-\alpha t} +.... )$$
$$={\alpha^{2}\over 24\pi G} + {\alpha^{2}\beta \over 12\pi G}\enskip a_{c}^{-3}
+{\alpha^{2}\beta\over 24\pi G}\enskip a_{c}^{-3}( e^{1\over (2n+1)} - 1 )$$
$$={\alpha^{2}\over 24\pi G} + {\alpha^{2}\beta \over 24\pi G}\enskip a_{c}^{-3}
+{\alpha^{2}\beta\over 24\pi G}\enskip a_{c}^{-3} e^{1\over (2n+1)}\eqno(38)$$
As $n\rightarrow\infty$, $e^{1\over(2n+1)}\rightarrow 1$
and $\rho_{c}\rightarrow \rho_{DE} +\rho_{DM}$  as before. Thus the assumptions leading to 
$(28)$ are consistent with the other inputs.

There are  five  parameters of the theory to start with, {\it viz.}, 
$\alpha, \beta, {\sqrt C}, A_{1}, A_{2}$,
with 
$\alpha \beta =24 \pi G {\sqrt C} A_{1}$.So we are left with 4 independent parameters.
We chose ${\sqrt C}={\alpha c'\over 24\pi G A_{1}}$ 
while determining the sound speed and 
this means $\beta=c'>> 1$ (discussion after equation $(33)$)
leaving 3 independent parameters,{\it viz.}, $\alpha, c'$ and $ A_{2}$. 
Now $A_{0}, A_{eq}, A_{m}$
are all positive numerical parameters and following the discussion after 
equation $(31)$ , $A_{0} > 1 $. In the present epoch $n\approx 2$, [1].
Hence $e^{\alpha t_{0}}\approx (5\beta)^{A_{0}}$. The scaling relation $(7)$ 
is valid provided $\Vert {V_{\phi}\over V}\Vert = {1\over e^{\alpha t}-\beta}\approx 0$. This means 
$e^{\alpha t} - \beta= 5^{A_{0}}\beta^{A_{0}}-\beta$ is very very large. 
But $A_{0}>1$ and $\beta=c'>> 1$. Therefore,
$e^{\alpha t} - \beta\sim 5^{A_{0}}\beta^{A_{0}}$ is indeed large and 
$(7)$ can be made to remain true by suitably choosing $A_{0},\beta$
to be very large.
Note that bounds on $\alpha$ may be estimated consistently from equations $(36)$ and 
$(21c)$ by using the current value of $H$ along with the values of 
$t_{0}, t_{eq}, t_{m}$ and suitably adjusting 
$A_{0}, A_{eq}, A_{m}$ and $c'$ (since $\beta=c'$).
So effectively we have a theory where there are two free parameters {\it viz.}, $c', A_{2}$,
to be adjusted consistently taking into account 
observations as far as possible. The value of  $A_{2}$ in $V(\phi)$ has to be consistently
chosen with the other parameters.Beyond this nothing more 
can be said regarding this constant.
 
{\bf 7. Discussion}

There is considerable literature on k-essence [14-25] and references therein. The present
work differs from all of these in that we have used a standard lagrangian with canonical
kinetic terms (obtained after a re-definition of variables) and used the solutions of the 
Euler -lagrange equations to directly determine cosmologically relevant quantities. Realistic cosmological scenarios are obtained in the context of dark matter and dark energy.
The basic  results can be summarised thus:
(a) $q(t)$ and $\phi$ can be looked upon as  dynamical 
variables whose classical time evolution can be obtained by solving  
the classical Euler-Lagrange equations corresponding to a 
lagrangian with canonical kinetic terms.
Assuming ${V^2\over {dV\over d\phi}}$ to be  a constant
gives a potential of the form 
$V(\phi)= {A_{1}\over \phi+A_{2}}$ where $A_{1},A_{2}$ are constants.
${{dV\over d\phi}\over V}$ can be made negligible 
so that the scaling relation of reference [1] can be 
made to remain valid. 
(b)The energy density $\rho$ has a dark energy component and a  
dark matter component.
(c)The pressure $p$
is {\it negative} for time $t\rightarrow \infty$ and
the sound velocity $c_{s}^{2}={\partial p\over\partial\rho} << 1$.
(d)The deceleration parameter $Q\rightarrow -1$ for time $t\rightarrow\infty$ 
and $Q\sim {1\over 2}$ for the matter dominated era.
(e)During the matter dominated era the equation of state parameter $w\approx 0$ and 
when the dark energy dominates (for times $t\rightarrow\infty$) we have $w\approx -1$.


\begin{thebibliography}{10}
\bibitem [1]{kn:xx}R.J.Scherrer, {\it Phys.Rev.Lett} {\bf 93} (2004), p.011301.
\bibitem [2]{kn:xx}C.Armendariz-Picon, T.Damour and V.Mukhanov, {\it Phys.Lett.} {\bf B458} (1999), p.209.
\bibitem [3]{kn:xx}J.Garriga and V.F.Mukhanov, {\it Phys.Lett.} {\bf B458} (1999), p.219.
\bibitem [4]{kn:xx}T.Chiba, T.Okabe and M.Yamaguchi {\it Phys.Rev.} {\bf D62} (2000), p.023511.
\bibitem [5]{kn:xx}C.Armendariz-Picon, V.Mukhanov and P.J.Steinhardt{\it Phys.Rev.Lett.} {\bf 85} (2000),p.4438.
\bibitem [6]{kn:xx}C.Armendariz-Picon, V.Mukhanov and P.J.Steinhardt {\it Phys.Rev.} {\bf D63} (2001),p.103510.
\bibitem [7]{kn:xx}T.Chiba, {\it Phys.Rev.} {\bf D66} (2002),p.063514.
\bibitem [8]{kn:xx}L.P.Chimento,{\it Phys.Rev.} {\bf D69} (2004),p.123517 ;
L.P.Chimento and A. Feinstein ,{\it Mod.Phys.Lett.} {\bf A19} (2004),p.761.
\bibitem [9]{kn:xx}M.Gasperini and G.Veneziano, {\it Phys.Rep.} {\bf 373} (2003), p.1.
\bibitem [10]{kn:xx}K.I.Maeda, {\it Phys.Rev.} {\bf D39} (1989),p.3159.
\bibitem [11]{kn:xx}V.Sahni, {\it Lect.Notes Phys.} {\bf 653} (2004), p.141.
\bibitem [12]{kn:xx}T.Padmanabhan, {\it AIP Conf.Proc.} {\bf 843} (2006), p.111.
\bibitem [13]{kn:xx}E.J.Kopeland,M.Sami and S.Tsujikawa, {\it Int.Jour.Mod.Phys.} {\bf D15} (2006), p.1753.
\bibitem [14]{kn:xx}M.Malaquarti,E.J.Copeland,A.R.Liddle and M.Trodden {\it Phys.Rev.} {\bf D67} (2003), p.123503 ;
M.Malaquarti,E.J.Copeland,A.R.Liddle,{\it Phys.Rev.} {\bf D68} (2003), p.023512.
\bibitem [15]{kn:xx}L.Mingzhe and X.Zhang, {\it Phys.Lett.} {\bf B573} (2003),p.20.
\bibitem [16]{kn:xx}J.M.Aquirregabiria,L.P.Chimento and R.Lazkoz, {\it Phys.Rev.} {\bf D70} (2004),p.023509.
\bibitem [17]{kn:xx}L.P.Chimento and R.Lazkoz, {\it Phys.Rev.} {\bf D71} (2005),p.023505.
\bibitem [18]{kn:xx}L.P.Chimento,M.Forte and R.Lazkoz, {\it Mod.Phys.Lett.} {\bf A20} (2005), p.2075.
\bibitem [19]{kn:xx}R.Lazkoz {\it Int.Jour.Mod.Phys.} {\bf D14} (2005),p.635.
\bibitem [20]{kn:xx}H.Kim {\it Phys.Lett.} {\bf B606} (2005), p.223.
\bibitem [21]{kn:xx}J.M.Aquirregabiria,L.P.Chimento and R.Lazkoz, {\it Phys.Lett.} {\bf B631} (2005), p.93.
\bibitem [22]{kn:xx}H.Wei and R.G.Cai, {\it Phys.Rev.} {\bf D71} (2005), p.043504.
\bibitem [23]{kn:xx}C.Armendariz-Picon and E.A.Lim, {\it JCAP} {\bf 0508} (2005), p.007.
\bibitem [24]{kn:xx}L.R.Abramo and N.Pinto-Neto {\it Phys.Rev.} {\bf D73} (2006),p.063522.
\bibitem [25]{kn:xx}A.D.Rendall, {\it Class.Quant.Grav.} {\bf 23}  (2006), p.1557.

\end{thebibliography}
\end{document}